\documentclass[11pt]{article}

\newcommand{\p}{\partial}

\newcommand{\aq}{\begin{eqnarray}}
\newcommand{\qa}{\end{eqnarray}}

\begin{document}

\begin{titlepage}

\begin{center}

{\Large \bf
{ Symmetries of the near horizon of a Black Hole by Group Theoretic methods }}

\vspace{10mm}

{\large{ K. Maharana } \\
\vskip 1.5cm
 {Physics Department, Utkal University, Bhubaneswar
  751 004, India}  \\

\sf karmadev@iopb.res.in  }

\vskip 3.5cm

\noindent
{{\bf{Abstract}}}
\end{center}

{ We use group theoretic methods to obtain the extended Lie point symmetries 
of the quantum dynamics  of a scalar
particle probing the near horizon structure of a black hole.
Symmetries of the classical equations of motion for a 
charged particle in the field of an inverse square potential and a
 monopole, in the presence of  certain 
model magnetic fields and potentials are also studied.    
Our analysis gives the generators and Lie algebras  generating  
the inherent symmetries. }

\end{titlepage}

\newpage

\section{Introduction}  

In certain physical problems there may exist extra hidden symmetries
which are not apparent, unless searched for.
Some of the examples from classical considerations are,
a particle in a $\frac{1}{r^2}$ potential in one dimension \cite{deA} 
and its quantum version in any dimension\cite{jackiw0}, two
dimensional
delta function potential in spatial two dimensions\cite{jackiw00},
 the conserved Runge-Lenz vector
of the Kepler problem and the extra symmetries of
a charge moving in the field of a magnetic 
monopole\cite{{jackiw},{jackiw1}}, and the
 generators beyond the Poincare invariance that give rise to
conformal invariance in electrodynamics as well as in Yang-Mills theory. 
The existence of symmetries
help in classifying and obtaining energy levels and eigenstates
in quantum mechanical problems, generating new solutions
and also formulating conservation laws. As we know,
manifestation of scale invariance in deep inelastic scattering
had deep significance in the development of gauge theories. The 
invariance under
scale and conformal transformations also motivated the
construction of a simple 
classical model which leads to  conformal quantum mechanics\cite{deA}.
Recently there have been a revival of interest in this model. This is
due to the observation
in string theory dynamics that a particle near a black hole possesses
SO(2,1) symmetries as in conformal quantum mechanics\cite{Sen}.

 The symmetries of charged particle - monopole system  and the 
conformal quantum mechanics were obtained from physical
reasonings and scale invariance in \cite{{deA},{jackiw}}. However, there
exists a general program  to obtain
 the symmetries of the equations of motion of any such system by
 using the group theoretic methods of Lie. 
In this paper we use this method to find the Lie point symmetries 
of the inverse squared potential and the monopole
system as well as some other physically motivated systems.

A knowledge of the symmetry group of a system of differential equations  
leads to several types of applications\cite{{olver},{ovs},{hill}} such as
finding the solutions,
constructing new solutions from the known ones . 
  Further, just the enumeration of the symmetry generators sometimes 
provide much physical insight and quantitative physical results for 
which the full solutions are not required. The derivation of Kepler's 
third law of planetary motion
and Runge-Lenz vectors\cite{stephani}, calculation of energy levels
for hydrogen like atoms and generalized Kepler's problems, harmonic 
oscillators,  Morse 
potentials\cite{wybourne}, electron in a specific nonuniform
 magnetic field\cite{mah03}, 
being some  such examples. In these the energy eigenvalues are 
obtained through the method of spectrum generating
 algebras which gives the 
Casimir invariants directly without explicit recourse to the solutions.
Of course, the solutions are also obtained from the representation
theory.
For finding the continuous symmetries, Lie's method of 
group analysis seems to be the most powerful technique available.
For example,  Witten\cite{witten}
had considered an example of the equation of motion of a
 particle in three
dimensions constrained to move on the surface of a sphere in 
the presence
of a magnetic monopole. This is the classical analogue of the 
Wess-Zumino
model\cite{wess}.  The equations of motion of such a system in the
 presence of a magnetic monopole cannot be obtained from the usual
 Lagrangian
formulation unless one goes   to a higher dimension. Hence, the usual
method of finding the symmetries through  Noether's
 theorem would have
difficulties. So, to look for the continuous symmetries associated
 with such classical
systems one has to analyze directly the equations of motion.
Similar is the case for Korteweg-de Vries equation which is not
 amenable to a
direct Lagrangian formulation  when expressed as a  lowest order 
equation\cite{blumen}.
Another example is the Lorenz system of equations which have been dealt
in the papers by Sen and Tabor\cite{sen}, and Nucci\cite{nuccilorenz}. 
For the classical systems,
 this procedure of finding directly the symmetries from 
equation of motion
is, in some sense, more fundamental. This is because
in certain cases many different Lagrangians may  give
 rise to the same
equations of motion. A group analysis of the equations of 
motion gives
all the Lie point generators of the symmetry  group. In the cases where 
a Lagrangian 
formulation is possible, the usual Noether symmetries are a subset of 
the above generators. This subset of generators acting on the
Lagrangian gives  zero\cite{stephani}.
However, besides these
there may be other generators obtained through group
analysis which have direct physical
significance, but not explicitly available from the consideration
of usual Noether symmetries alone\cite{{prince},{BC},{NC},{krause}}. The reproduction
of Kepler's third law in the planetary motion problem 
is such an example.
 The extension of this
idea to the notion of Lie dynamical symmetries contains
 similarly a subclass
known as Cartan symmetries. The Runge-Lenz vector can be 
obtained from
such considerations. These symmetries are, further, related
 to the Lie-B\"{a}cklund symmetries.

The application of these types of analysis to nonlocal cases have been
widely studied through B\"{a}cklund transformations and related
techniques in the context of integrable systems containing infinite
number of conservation laws\cite{ibragimov}. In \cite{Leo} a
hereditary 
recursion operator for the Harry-Dym equation generating infinitely many 
Lie-B\"acklund symmetries have been found. The nonabelian  prolongation 
structure of the 
cylindrical Korteweg-de Vries has been used to derive a set of B\"acklund
transformations and a nonlinear superposition formula in \cite{Leo1}.
 The Thirring model
 has been analyzed by
Morris \cite{{morris},{Dav}}. The differential geometric forms 
developed earlier
 are used in the above analysis to obtain the 
prolongation structure\cite{estabrook}.
 Some other applications  of these ideas to important problems from 
physics is comprehensibly covered by Gaeta\cite{gaeta}.  

In this paper, we  
find the Lie point
symmetries of  equations 
representing the motion of a charged particle in three dimensions
under the influence of (i) a $\frac{1}{r^2}$ potential, (ii) in field of a
monopole, (iii) classical Wess-Zumino-Witten model, (iv) a dyon, and,  
(v) a scalar field probing the near horizon structure 
of a black hole. 

\section{ Symmetry generators, classical particle }

 Typically we are interested in the coupled set
of equations representing the equations of motion of a particle in
three dimensions. These are of the form
\begin{eqnarray}
  {{\ddot{x}}_{a}} = {\beta}{{\omega}_{a}} {({{x_{i}}},{\dot{x}}_{i},t )} 
   \label{eq:nfree}
\end{eqnarray}
where a dot represents derivative with respect to time, $ {a,i = 1,2,}$
and  $3$, and $ {\beta} $ is a constant involving mass,
coupling constant etc. which we set equal to one. 
Following Stephani \cite{stephani}, we will find the infinitesimal 
generators of the symmetry under which the system of
differential equations does not change.  The 
symmetry is generated by $\bf{X} $ and its extension
\begin{eqnarray}
   {\dot{\bf{X}}} &=& \tau {\frac{\p}{\p{ t}}} + 
   {{\eta}_a} {\frac{\p}{{\p}{x_a}}} + 
   {\dot{\eta}}_a {\frac{\p}{{\p{\dot{x}}_a}}}  
\end{eqnarray}
and the symmetry condition under transformations
represented by the following equation  determines $\tau$ and $ {\eta}_a$s.
In the expanded form the 
symmetry conditions are given by
\begin{eqnarray}
    {\eta}_b {{\omega}_a}_{,b} + {( {{\eta}_b}_{,t}
   + {\dot{x}}_c {{\eta}_b}_{,c} - {\dot{x}}_b  {{\tau}_{,t}}
   -  {\dot{x}}_b {\dot{x}}_c  {{\tau}_{,c}} )}
   { \frac{\p{{\omega}_a}}{\p{\dot{x}}_b}}  \nonumber \\ 
   +  {\tau} {{{\omega}_a}_{,t}} + 2 {{\omega}_a} {( {{\tau}_{,t} + 
   {\dot{x}}_b}{{\tau}_{,b}} )}  
   + {\omega_b}{({\dot{x}}_a {{\tau}_{,b}} - {{\eta}_a}_{,b}) }  \nonumber \\
   + {\dot{x}}_a  {\dot{x}}_b  {\dot{x}}_c  {\tau}_{,bc} + 
   {\dot{x}}_a {{\tau}_{,tt}}  
   + 2 {\dot{x}}_a  {\dot{x}}_c {{\tau}_{,tc}}   \nonumber \\ 
    -  {\dot{x}}_c  {\dot{x}}_b  {{\eta}_a}_{,bc}
   - 2 {\dot{x}}_b {{\eta}_a}_{,tb}
   - {{{\eta}_a}_{,tt}}  = 0  \label{eq:condition}
\end{eqnarray}
where $ f_{,t} = {\frac{{\p}f}{\p{t}}} $ and $ f_{,c} 
  = {\frac{{\p}f}{\p{x_c}}} $.

The solutions of the symmetry conditions provide us the generators of the 
group.

 In the presence of a potential like  $ \frac{1}{r^2} $ we find
 the generators with extensions to be
\aq
{\dot{\bf {X}}_a } =  {{\varepsilon}_{abc}} \left( {x_c}\frac{\p}{\p x_b}
   + {{\dot x}_c {\frac{\p}{\p {\dot x}_b}}} \right), 
  \quad    space \ rotations,  \nonumber \\
 \dot{{\bf {X}}_4 } =  { \frac{\p}{\p t}}, \quad time \ translation,  \nonumber \\
  {\dot{\bf {X}}_5 }  = 2 t  \frac{\p}{\p t}  + {x_a}\frac{\p}{\p x_a} -
  {\frac{1}{2}}  {{\dot x}_a}
 {\frac{\p}{\p {\dot x}_a}},     \nonumber \\ 
   \  Kepler \ like \ scaling \ law
 \ {\frac{t}{r^2}} = constant, \nonumber  \\
  {\dot{\bf {X}}_6 }  =  t^2  \frac{\p}{\p t}  + t {x_a}\frac{\p}{\p x_a} +
 {{ x}_a}
  {\frac{\p}{\p {\dot x}_a}}  -  {{\dot x}_a}
  {\frac{\p}{\p {\dot x}_a}}       
  \label{eq:mono}
\qa
The vector fields have the commutation relations
\aq
  \left[{\bf X }_{a}, {\bf X }_{b} \right] =
   {{\varepsilon}_{abc}} {\bf X }_{b},  \nonumber \\ 
   \left[{\bf X }_{a}, {{\bf X }_4} \right] = 
 \left[{\bf X }_{a}, {{\bf X }_5} \right] = 
 \left[{\bf X }_{a}, {{\bf X }_6} \right] =  0 \nonumber  \\
 \left[{\bf X }_{4}, {\bf X }_{5} \right] = 2 {\bf X }_{4},\
     \left[  {\bf X }_{4}, {\bf X }_{6} \right] =  {\bf X }_{5},\
   \left[  {\bf X }_{5}, {{\bf X }_{6}} \right] = 2 {{\bf X }_{6}}
\qa

The classical Kepler problem with $ \frac{1}{r} $ potential
has a different scaling law of $ \frac{t^2}{r^3} $ and also 
does not possess the symmetry corresponding to generator
${\bf X_6}$. However, it possesses a Runge-Lenz vector.
Stephani has given a general method to obtain such conserved
vectors in the Lagrangian formulation. In the quantum mechanical case,
if the eigenvalues are taken instead the Hamiltonian operator,
an enhanced symmetry occurs for  $ \frac{1}{r} $ potential.
For  $ \frac{1}{r^2} $ potential we could not find a classical Runge-Lenz
vector by Stephani's method. This appears to be related  to
orbits being not closed in such a potential\cite{{gold},{Gold},{Gold1}}. 
However, as 
has already been noted,
in this case new vector fields result leading to the  extra symmetries. 

For the quantum mechanical case with a  $ \frac{1}{r^2} $ potential
in the Schr\"odinger equation, Jackiw has given the
 physical argument that in any dimensions the kinetic term scales as
$ \frac{1}{r^2} $ and an $SO(2,1)$ symmetry results\cite{jackiw0}.
Jackiw has also considered the symmetries of equation of motion, Lagrangian,
and Hamiltonian for a charged particle in the field of a magnetic 
monopole \cite{{jackiw},{jackiw2},{DH}}. He had discovered an extra
  $SO(2,1)$ hidden symmetry by scaling and physical considerations.
Leonhardt and Piwnicki have explored the
theoretical possibility of obtaining the  field of
quantized monopoles when a classical dielectric moves in a charged
capacitor \cite{leonhardt}. 
Since the magnetic field due to a magnetic monopole is $ {B_a} = {\frac{x_a} 
{r^3}}  $, the equation of motion is
\aq
{{\ddot{x}}_a } = {{\varepsilon}_{abc}}  {\frac{{{\dot{x}}_b} {x_c}}
  {r^3}}  = {\omega_a} \label{eq:monopl}
\qa
The Lie symmetries of this equation were obtained in\cite{{Mor},{Haas}}
and are identical to those of the  $ \frac{1}{r^2} $ potential
found by us. Such a potential was  added by Zwanzinger in his analysis
of monopole system\cite{Zwa}.

\section{ Symmetry generators, quantum particle }

The simplest one dimensional version  of the equation
\aq
{{\ddot{x}}_a } =  {\frac{{ \mu}^2  { x_a}}{mr^4}}  \label{eq:fub}
\qa 
possesses remarkable symmetries which were exploited by de Alfaro,
 Fubini, and Furlan to construct conformal quantum mechanics. Here $x$ is 
considered as a field in
zero space and one time dimension.
The quantum mechanical equation for the wave function $u$
 becomes\cite{deA}, in our notation, 
\aq
  {\left( - {\frac{d^2}{{dr}^2}} + {\frac{g}{r^2}} +
 {\frac {r^2}{a^2}} \right)} u  =  {\frac{4b}{a}} u \label{eq:fub2}
\qa
where $   {\frac{{ \mu}^2 }{m}} $ is replaced by $g$. Here $a$ is a
constant which plays a fundamental role in the theory and $b$ is related
to appropriate raising and lowering operators. This equation
 can be expressed in terms of  differential operator realization
of $su(1,1)$ algebra\cite{wybourne} and was studied in detail 
 in\cite{deA}.

There have been earlier works, where it has been shown that the
 dynamics of a scalar particle approaching the event horizon of 
a black hole is governed by an Hamiltonian with an inverse 
square potential\cite{{berg},{Cl},{Gib},{Wt},{t1},{t2},{Sen}}. The scalar
 field can be 
used as a probe to study the geometry in the vicinity of the horizon
 and its dynamics is expected to provide clues to the inherent 
symmetry properties of the system.
The Hamiltonian of conformal quantum mechanics fits into this. This 
Hamiltonian also arises as a limiting case of the brick-wall model 
describing the low energy quantum dynamics of a field in the background 
of a massive Schwarzschild black hole of mass $M$ \cite{{t1},{t2}}. On 
factorizing such a Hamiltonian a Virasoro symmetry was found by 
Birmingham, Gupta and Sen. They
 have studied the representation of the algebra as well as the scaling 
properties of the time independent modes\cite{Sen}.
The full Virasoro algebra was obtained by the requirement of unitarity
of the representation. The Hamiltonian operator is in the enveloping algebra. 
  
However, here we aim at finding the underlying Lie point 
symmetry 
of the equation of motion of the scalar particle, viewed as a 
differential equation. 
 Of course, mathematically 
any two linear homogeneous
ordinary differential equations can be transformed to the form, where a 
prime denotes a differentiation with respect to
$r$,
\aq
   u''  = 0.
\qa
 This equation has the eight dimensional symmetry of projective 
transformations.  But the two equations could be quite different 
from physics 
point of view having different eigenvalues and eigenfunctions. Hence
 we would like to see explicitly what are the Lie point symmetries of the
particular equation.

For the equation
\begin{eqnarray}
  u'' = \omega (r,u,u') \label{eq:our}
\end{eqnarray}
where
\begin{eqnarray}
{\omega} (r,u,u') = - { ( {\frac{C}{r^2}} + {\frac{D}{r}}
  + \hat{E})} u(r).\label{eq:CDE}
\end{eqnarray}
the symmetry generators are obtained from the conditions given by the
equation  (\ref{eq:condition}) which reduces in the one dimensional case to
\begin{eqnarray}
\omega ( {\eta}_{,u} - 2 {\tau}_{,r} -3 u' {\tau}_{,u} )
 - {\omega}_{,r} \tau - {\omega}_{,u} {\eta}   \nonumber     \\
- {\omega}_{,{u'}} {[ {\eta}_{,r} + u' ( {\eta}_{,u} -{\tau}_{,r} )
 - {u'}^2 {\tau}_{,u} ]}    
+ {\eta}_{,rr}     \nonumber     \\  +    u' ( 2 {\eta}_{,ru} - {\tau}_{,rr} )
+ {u'}^2 ( {\eta}_{,uu} - 2 {\tau}_{,ru} ) - {u'}^3 {\tau}_{,uu}
= 0  \label{eq:oned}
\end{eqnarray}
For the case 
$C = - g $, $ D = \hat{E}  = 0 $,
equating to zero the coefficients of ${u'}^3 $ and ${u'}^2 $ in
(\ref{eq:oned})
we get
\begin{eqnarray}
{\tau}_{,uu} =0 , \qquad {\eta}_{,uu} = 2 {\tau}_{,ru}
\end{eqnarray}
which are satisfied for
\begin{eqnarray}
{\tau} = u \alpha (r) + \beta (r) , \quad  \eta  = u^2  {\alpha}' (r) +
 u {\gamma}(r) + \sigma (r)  .
\end{eqnarray}
Using these and equating to zero the coefficient of $u'$ 
 and then considering the 
 the terms not involving  $u'$, we find that
an interesting symmetry exists only when the coupling constant $g$ 
is equal to 2. For this case
 we obtain 
\aq
 {\tau} = {\frac{1}{r}} Au + Fr, \quad 
{\eta} =  -  {\frac{1}{r^2}} Au^2 + B u + {\sigma}(r) \label{eq:sen}
\qa
where $ A, F$, and $B$ are constants and ${\sigma}(r)$ satisfies 
the same equation as $u$ does. The vector fields are
\begin{eqnarray}
{{\bf X}_1} = r {\frac{\p}{\p r}},
  \quad {\bf X_2} = u {\frac{\p}{\p u}},  \quad
 {{\bf X}_3} =  {\frac{1}{r}}u {\frac{\p}{\p r}}
  -  {\frac{1}{r^2}}u^2 {\frac{\p}{\p u}} \label{eq:sn}
\end{eqnarray}
with commutation relations
\aq
[{{\bf  X}_1 } ,{ {\bf X}_2 } ] = 0, \quad 
 [{{\bf  X}_1 } , {{\bf X}_3 } ] =  -2  {{\bf X}_3 }, \quad  
[{{\bf  X}_2 } , {{\bf X}_3 } ] =  {{\bf X}_3 }
\qa
For the  case,  considered in \cite{Sen2}, 
  the relevant equation is
\begin{eqnarray}
 {\frac{d^2 u}{{dr}^2}}  + {\frac{1}{r^2}} 
 {\left[{\frac{1}{4} + {R^2}{E^2}}  \right]} u = 0 \label{S0}
\end{eqnarray}
where $E$ is a generic eigenvalue and $R = 2M$.
Hence $g$ corresponds to ${\left[{\frac{1}{4} + {R^2}{E^2}}\right]} $
 in this case.
This equation corresponds to the
Hamiltonian of the form
\aq
  H =  \frac{p^2}{2} +  \frac{ g'}{2r^2}
\qa
where $ - \frac{ g'}{2}   =  
     {\left[{\frac{1}{4} + {R^2}{E^2}}  \right]}$.
Along with $H$, the generators of dilations
\aq
  D =  \frac{(pr + rp)}{2},
\qa
and special conformal transformations
\aq
  K = \frac{r^2}{2}
\qa
form the $SL(2,R)$ conformal algebra.  But since the $D$ and $K$ do 
not commute with the Hamiltonian, as is well known,  $SL(2,R)$ is 
not a symmetry of the 
theory\cite{strom}. However,
 $SL(2,R)$ can be used to relate states of different energies.
Our analysis of the eigenvalue equation  shows, in contradistinction to the 
above formulation, what  the symmetries are. However,
  only when $E$ is imaginary
with  ${\left[{\frac{1}{4} + {R^2}{E^2}}\right]} = -2 $, the symmetry
will show up. This reminds of the situation in spontaneous symmetry breaking
in gauge theories where the mass squared value has to become negative.
But in our case, on the other hand, the symmetry pops up at 
the particular imaginary $E$.
The one parameter groups $G_i$ generated by the vector fields ${\bf X}_i $
are given by,
\aq
  G_1 : ({e^{\epsilon }r, u}), \nonumber  \\ 
  G_2 : (r ,{e^{\epsilon } u}), \nonumber  \\ 
  G_3 : { \left(  ( 1 +  \frac{\epsilon u}{r}
      - {\frac{{\epsilon}^2 u^2}{r^4}})  r, 
\frac{ exp{\{ - \frac{\epsilon u}{r^2}\}}}
    {(1 - \frac{{\epsilon}^2 u^2}{r^2 })} u  \right)}. 
\qa
 up to order ${\epsilon}^2$ for $G_3$.
The wave function exhibits scaling behaviour at this imaginary $E$
for the generators ${\bf{X}_2}$  and shows a much complicated symmetry 
behaviour corresponding to nonlinear transformation involving
$u$ and $r$ for the vectorfield ${\bf{X}_3}$.

 We note here that the $ {\frac{1}{r}}$
and $ {\frac{1}{r^2}}$ factors in ${{\bf X}_3}$ makes it ill defined as
$r \rightarrow 0 $ similar to the $L_{-n}$ operators of conformal field theory 
or the $ P_m $ operators considered by Birmingham, Gupta, and Sen\cite{Sen}.

\section{Conclusion}

The equations of 
motion of a free particle, admit eight symmetries for 
each of the ${x_a}$s. This
is the maximum number of symmetries for an
ordinary second order differential equation. By including different
$ {x_a} $, $ {\dot{x_a}} $ dependent terms in the equations we
do explicitly see which generators survive as symmetries and we have found
corresponding  complete Lie algebras. We choose
some cases motivated by problems from
physics. The original motivation of including Wess-Zumino terms
in the Lagrangian has been to reduce some of its symmetries\cite{witten2},
By above type of analysis we 
 find that  the equations of motion now support the three dimensional
 rotations and a time translation symmetry instead of the 
six vector fields as for the
 monopole problem without any constraint. A calculation gives 
the same four symmetries  for 
dyon. It should be noted that in the context of the symmetries of 
Wess-Zumino-Witten
 models, the  symmetries in the higher dimensions play by far
the most important role and these have been fruitfully exploited  
 \cite{{braaten},{For}}.  

  The Lie symmetries correspond to the transformations of
the solutions. An analysis for Dirac equation modified by nonlinearity
has been carried out in \cite{mahE}.
It is also expected that related group  analysis may provide useful
information
when terms are modified in the Lagrangian, due to quantum corrections,
for example.

For the case of a scalar particle probing the near horizon structure of a
black hole, under certain limits the Hamiltonian contains $\frac{1}{r}$
and $\frac{1}{r^2} $ potentials.  Here we find that only for a specific value
of the coefficient of $\frac{1}{r^2}$ term and in 
absence of $\frac{1}{r}$ term  there exist  symmetry with
three generators. The symmetries are a scaling transformation of $r$, 
a scaling transformation of the wave function $u$, and 
a nonlinear transformation
involving both $u$ and $r$ for $u$.  The specific value of energy
 corresponds to
energy squared being negative, whereas for spontaneous symmetry breaking to
 occur in gauge theories, the mass squared is taken  to be negative.
We note that in contrast to spontaneous  breaking of symmetry, 
our analysis shows that the symmetry of the eigenvalue equation
is enhanced for a particular imaginary value of the energy.  
We expect that these considerations will ultimately lead to a better
understanding of the spontaneous symmetry breaking, as well.

The similar  symmetries of an inverse square potential and 
the monopole system in 
the classical case is intriguing. Symmetry analysis
of the monopole system   for the quantum case would lead to a deeper 
understanding.
The quantum mechanical problem of a charged particle
in the presence of even a constant magnetic field has many interesting 
mathematical structures \cite{{plyus},{plyus1}} and under certain 
limits can make 
space coordinates noncommutative \cite{bigatti}.  Klishevich and Plyuschay 
have found a universal algebraic 
structure at the quantum level for the two dimensional case in the
presence of certain magnetic fields\cite{{plyus},{plyus1}}. 
Nonlinear superconformal
symmetry of the fermion-monopole system has been 
extensively studied in\cite{{plyus2},{plyus3}}.
It would be worthwhile to analyze  the existence of such structures 
in combination with various $r$ dependent potentials.

\noindent{\bf Acknowledgement}

{I  thank Professor Dieter L\"ust,  Arnold Sommerfeld Centre, 
Ludwig Maximilians
University, Munich, and Professor Hermann Nicolai,  Albert Einstein 
Institute, Max Planck Institute for
Gravitational Physics, Golm, for providing  warm
hospitality and  academic
facilities  to carry out this work. I am grateful to Professor Roman
Jackiw for an instructive communication and to the referee for
constructive comments. }




\end{document}